# The 2005 Neyman Lecture: Dynamic Indeterminism in Science[1]

**David R. Brillinger**

*Abstract.* Jerzy Neyman's life history and some of his contributions to applied statistics are reviewed. In a 1960 article he wrote: "*Currently in the period of dynamic indeterminism in science, there is hardly a serious piece of research which, if treated realistically, does not involve operations on stochastic processes. The time has arrived for the theory of stochastic processes to become an item of usual equipment of every applied statistician.*" The emphasis in this article is on stochastic processes and on stochastic process data analysis. A number of data sets and corresponding substantive questions are addressed. The data sets concern sardine depletion, blowfly dynamics, weather modification, elk movement and seal journeying. Three of the examples are from Neyman's work and four from the author's joint work with collaborators.

## 1. INTRODUCTION

This paper is meant to be a tribute to Jerzy Neyman's substantive work with data sets. There is an emphasis on scientific questions, statistical modeling and inference for stochastic processes.

The title of this work comes from Neyman (1960) where one finds,

> "The essence of dynamic indeterminism in science consists in an effort to invent a hypothetical chance mechanism, called a 'stochastic model,' operating on various clearly defined hypothetical entities, such that the resulting frequencies of various possible outcomes correspond approximately to those actually observed."

Here and elsewhere Neyman appeared to use the adjective "indeterministic" where others would use "stochastic," "statistical" or "nondeterministic"; see, for example, Neyman and Scott (1959). Perhaps Neyman had some deeper or historical context in mind, but that is not clear. In this paper the emphasis is on the word "dynamic."

Jerzy Neyman (JN) led a full life. Reid (1998) contains many details and anecdotes, a lot of them in Neyman's own words. Other sources include the papers: Neyman (1970), Le Cam and Lehmann (1974), Kendall, Bartlett and Page (1982), Scott (1985), Lehmann (1994) and Le Cam (1995).

The article has six sections: 1. Introduction, 2. Jerzy Neyman, 3. Some formal methods, 4. Three examples of JN's applied statistics work, 5. Four examples of random process data analysis, 6. Conclusion. The focus is on applied work in the environmental sciences and phenomena. This last is a word that Neyman often employed.

*David Brillinger is Professor, Department of Statistics, University of California, Berkeley, California 94720, USA (e-mail: brill@stat.berkeley.edu).*

[1]Discussed in 10.1214/07-STS246B and 10.1214/07-STS246A; rejoinder at 10.1214/07-STS246REJ.

In particular the examples show how random process modeling can prove both helpful and not all that difficult to implement. The thought driving this paper is that by examining a number of examples, unifying methods and principles may become apparent. One connecting thread is "synthetic" data, in the language of Neyman, Scott and Shane (1953) and Neyman and Scott (1956). Synthetic data, based on simulations, are an exploratory tool for model validation that has the advantage of suggesting how to create another model if the resemblance of the simulation to the actual data is not good.

There are quotes throughout to create a flavor of JN's statistical approaches.

## 2. JERZY NEYMAN

> "His devotion to Poland and its culture and traditions was very marked, and when his influence on statistics and statisticians had become worldwide it was fashionable ... to say that 'we have all learned to speak statistics with a Polish accent' ..." (Kendall, Bartlett and Page, 1982).

The life of Neyman is well documented by JN and others; see, for example, Reid (1998), LeCam and Lehmann (1974) and Scott (1985). Other sources are cited later. Neyman was of Polish ancestry and as the above quote makes clear he was very Polish! Table 1 records some of the basic events of his life. One sees a flow from Poland to London to Berkeley with many sidetrips intermingled throughout his life. These details are from Scott (1985) and Reid (1998).

Neyman's education involved a lot of formal mathematics (integration, analysis, ...) and probability. He often mentioned the book, *The Grammar of Science* (Pearson, 1900) as having been very important for his scientific and statistical work. He described Lebesgue's *Leçons sur l'intégration* as "the most beautiful monograph that I ever read."

The Author's Note to the *Early Statistical Papers* (Neyman, 1967) comments on the famous and influential teachers he had at Kharkov. They included S. Bernstein ("my teacher in probability"), C. K. Russyan, and A. Przeborski. Others he mentions as influential include E. Borel, R. von Mises, A. N. Kolmogorov, E. S. Pearson and R. A. Fisher.

Neyman came to Berkeley in 1938. That appointment had been preceded by a triumphant U.S. tour in 1937. The book Neyman (1938b) resulted from the tour. After Neyman's arrival, internationally renowned probabilists and statisticians began to visit Berkeley regularly and contributed much to its research atmosphere and work ethic.

In Neyman's time the lunch room used to play an important role in the Berkeley Department. JN, Betty Scott (ELS) and Lucien Le Cam enthralled students, colleagues, visitors and the like with their conversation. They involved everyone in the stories and discussions.

Neyman had a seminar Wednesday afternoons. It began with coffee and cakes. Then there was a talk, often by a substantive scientist, but theoretical talks did occur from time to time. The talk's discussion was followed by drinks at the Faculty Club including the famous Neyman toasts. "To the speaker. To the international intellectual community. To the ladies present and some ladies absent." Up until perhaps the mid-1970s there was a dinner to end the event.

Neyman's work ethic was very strong. It typically included Saturdays in the Department, and for those who came to work also there were cakes at 3 pm.

## 3. SOME FORMAL METHODS

> "Every attempt at a mathematical treatment of phenomena must begin by building a simplified mathematical model of the phenomena." (Neyman, 1947).

This section provides a few of the technical ideas and methods that are basic to the examples pre-sented. The examples involve dynamics, time, spatial movement, Markov processes, state-space models, stochastic differential equations (SDEs) and phenomena.

Table 1
*A timeline of Jerzy Neyman's life*

| Date | Event |
| --- | --- |
| 1894 | Born, Bendery, Monrovia |
| 1916 | Candidate in Mathematics, University of Kharkov |
| 1917–1921 | Lecturer, Institute of Technology, Kharkov |
| 1921–1923 | Statistician, Agricultural Research Institute, Bydgoszcz, Poland |
| 1923 | Ph.D. in Mathematics, University of Warsaw |
| 1923–1934 | Lecturer, University of Warsaw |
|  | Head, Biometric Laboratory, Nencki Institute |
| 1934–1938 | Lecturer, then Reader, University College, London |
| 1938–1961 | Professor, University of California, Berkeley |
| 1955 | Berkeley Statistics Department formed |
| 1961–1981 | Professor Emeritus, University of California, Berkeley |
| 1981 | Died, Oakland, California |

### 3.1 Random Process Methods

> "..., modern science and technology provide statistical problems with observable random variables taking their values in functional spaces." (Neyman, 1966).

By a random process is meant a random function. Their importance was already referred to in Section 1. In particular Neyman was concerned with "phenomena developing in time and space" (Neyman, 1960). The random processes describing these are the backbone of much of modern science.

### 3.2 Markov Processes

Neyman was taken with Markov processes. Reid (1998) quotes him as saying,

> "So what Markov did—he considered changes from one position to another position. A simple example. You consider a particle. It's maybe human. And it can be in any number of states. And this set of states may be finite, may be infinite. Now when it's Markov—Markov is when the probability of going—let's say—between today and tomorrow, whatever, depends only on where you are today. That's Markovian. If it depends on something that happened yesterday, or before yesterday, that is a generalization of Markovian."

Time and Markovs play key roles in Fix and Neyman (1951). An advantage of working with a Markov process is that when there is a parameter one can set down a likelihood function directly.

### 3.3 Stochastic Differential Equations (SDEs)

> "It seems to me that the proper way of approaching economic problems mathematically is by equations of the above type, in finite or infinitesimal differences, with coefficients that are not constants, but random variables; or what is called random or stochastic equations. ... The theory of random differential and other equations, and the theory of random curves, are just starting." (Neyman, 1938a).

To give an example, let $\mathbf{r}(t)$ refer to the location of a particle at time $t$ in $R^p$ space. The path that it maps out as $t$ increases is called the trajectory. (Trajectory is an old word used for a stochastic process.) Its vector-valued velocity will be denoted

$$\boldsymbol{\mu}(t) = d\mathbf{r}(t)/dt.$$

Rewriting this equation in terms of increments and adding a random disturbance leads to a so-called stochastic differential equation

$$(1) \qquad d\mathbf{r}(t) = \boldsymbol{\mu}(\mathbf{r}(t),t)\,dt + \boldsymbol{\sigma}(\mathbf{r}(t),t)\,d\mathbf{B}(t)$$

or in integrated form,

$$(2)\ \mathbf{r}(t) = \mathbf{r}(0) + \int_0^t \boldsymbol{\mu}(\mathbf{r}(s),s)\,ds + \int_0^t \boldsymbol{\sigma}(\mathbf{r},s)\,d\mathbf{B}(s).$$

If, for example, the process $\mathbf{B}$ is Brownian, that is, the increments $\mathbf{B}(t_{i+1}) - \mathbf{B}(t_i)$ are $IN(\mathbf{O}, (t_{i+1} - t_i)\mathbf{I})$, then, under conditions on $\boldsymbol{\mu}$ and $\boldsymbol{\sigma}$, a solution of the equation exists and is a Markov process. The function $\boldsymbol{\mu}$ is called the drift rate and $\boldsymbol{\sigma}$ the diffusion coefficient.

A particular case of an SDE is the Ornstein–Uhlenbeck process given by

$$d\mathbf{r}(t) = \alpha(\mathbf{a} - \mathbf{r(t)})\,dt + \sigma\,d\mathbf{B}(t)$$

with $\alpha > 0$ and $\sigma$ a scalar. This models a particle being attracted to the point $\mathbf{a}$ with the motion disturbed randomly.

An approximate solution to (1) is given, recursively, by

$$\begin{aligned}(3)\quad \mathbf{r}(t_{i+1}) - \mathbf{r}(t_i) &\approx \boldsymbol{\mu}(\mathbf{r}(t_i),t_i)(t_{i+1} - t_i) \\ &\quad + \boldsymbol{\sigma}(\mathbf{r}(t_i),t_i)\mathbf{Z}_i\sqrt{t_{i+1}-t_i}\end{aligned}$$

with the $t_i$ an increasing sequence of time points filling in the time domain of the problem; see Kloeden and Platen (1995). The $\mathbf{Z}_i$ are independent $p$-variate standard normals. This solution procedure to (1) is known as the Euler method. In fact Itô (1951) used an expression like (3) to demonstrate that, under conditions, (1) had a unique solution.

There has been a substantial amount of work on statistical inference for SDEs; references include Heyde (1994) and Sørensen (1997). There are parametric and nonparametric fitting methods. Inferential work may be motivated by setting down the above approximation and taking the $t_i$ to be the times of observation of the process.

Assuming that $\boldsymbol{\mu}(\mathbf{r}, t) = \boldsymbol{\mu}(\mathbf{r})$, that $\boldsymbol{\sigma}(\mathbf{r}(t), t) = \sigma \mathbf{I}$, $\sigma$ scalar, and that $\mathbf{r}$ is $p$ vector-valued, one can consider as an estimate of $\sigma^2$

$$\hat{\sigma}^2 = \frac{1}{pI} \sum_i \|\mathbf{r}(t_{i+1}) - \mathbf{r}(t_i) - \hat{\boldsymbol{\mu}}(\mathbf{r}(t_i))$$
(4)
$$\cdot (t_{i+1} - t_i)\|^2 / (t_{i+1} - t_i),$$

$i = 1, \ldots, I$, having determined an estimate of $\boldsymbol{\mu}$.

If the region of motion, say $D$, is bounded with boundary $\partial D$, one can proceed via the SDE

$$d\mathbf{r}(t) = \boldsymbol{\mu}(\mathbf{r}(t), t) dt + \boldsymbol{\sigma}(\mathbf{r}(t), t) d\mathbf{B}(t) + d\mathbf{A}(t)$$

with the support of $\mathbf{A}$ on the boundary $\partial D$. This construction pushes the particle into $D$.

### 3.4 A Potential Function Approach

The choice of the function $\boldsymbol{\mu}$ in (1) may be motivated by Newtonian dynamics. Suppose there is a scalar-valued potential function, $H(\mathbf{r}(t), t)$; see Taylor (2005). Such a function $H$ can control a particle's direction and velocity.

In a particular physical situation the Newtonian equations of motion may take the form

$$d\mathbf{r}(t) = \mathbf{v}(t) dt,$$
(5) $\quad d\mathbf{v}(t) = -\beta \mathbf{v}(t) dt - \beta \nabla H(\mathbf{r}(t), t) dt,$

with $\mathbf{r}(t)$ the particle's location at time t, $\mathbf{v}(t)$ the particle's velocity and $-\beta \nabla H$ the external force field acting on the particle. The parameter $\beta$ represents the coefficient of friction. Here $\nabla = (\partial/\partial x, \partial/\partial y)^\tau$ is the gradient operator. For example, Nelson (1967) makes use of the form (5).

In the case that the relaxation time, $\beta^{-1}$, is small (or in other words, the friction is high), (5) is approximately

$$d\mathbf{r}(t) = -\nabla H(\mathbf{r}(t), t) dt = \boldsymbol{\mu}(\mathbf{r}, t) dt.$$

Writing the velocity $\mathbf{v}(t) = \boldsymbol{\mu}(\mathbf{r}, t)$ one is led to a stochastic gradient system

$$d\mathbf{r}(t) = -\nabla H(\mathbf{r}(t), t) dt + \sigma d\mathbf{B}(t).$$

The function $H$ might be a linear combination of elementary known functions, a combination of thin plate splines placed around a regular grid or based on a kernel function. Example 7 below will indicate the method. The method is further elaborated in Brillinger (2007a, 2007b).

## 4. THREE EXAMPLES OF JN'S APPLIED STATISTICS WORK

> "...the delight I experience in trying to fathom the chance mechanisms of phenomena in the empirical world." (Neyman, 1970).

Neyman was both an exceptional mathematical statistician and an exceptional applied statistician. The applied work commenced right at the beginning of his career and continued until the very end. This section presents examples from astronomy, fisheries and weather modification. These examples were chosen as they are interesting and they blend into the later examples in the paper.

Neyman's work was special in applied statistics in that he set down specific "postulates" or assumptions. Tools of his applied work included sampling, best asympotically normal (BAN) estimators, C($\alpha$) tests, chi-squared, randomization and synthetic data. His work was further characterized by the very careful preparation of the data by his Statistical Laboratory workers.

JN's applied papers typically include substantial introductions to the scientific field of concern. Topics include farfield effects of cloud seeding, estimation of the dispersion of the redshift of galaxies, higher-order clustering of galaxies, and sardine depletion.

Given Neyman's concern with the scientific method, one can wonder how he validated or appraised his models. On reading his papers, hypothesis testing seems to include assessment. There were lots of data, and fit components (observed–expected) and chi-squared (residuals). There was smooth chi-squared to get alternative hypotheses. There was often the remark, "appears reasonable."

### 4.1 Example 1. ASTRONOMY

> "By far the strongest and most sustained effort expended for us in studying natural phenomena through appropriately selected aspects of the process of clustering referred to astronomy, specifically to galaxies. ..., the stimulus came from the substantive scientists, that is from astronomers." (Neyman and Scott, 1972).

The work of Neyman, and his collaborators in this case, is a model for applied statistics. The question is made clear. Substantive science is involved. Statistical theory is employed and developed as necessary. Empirical analyses are carried out.

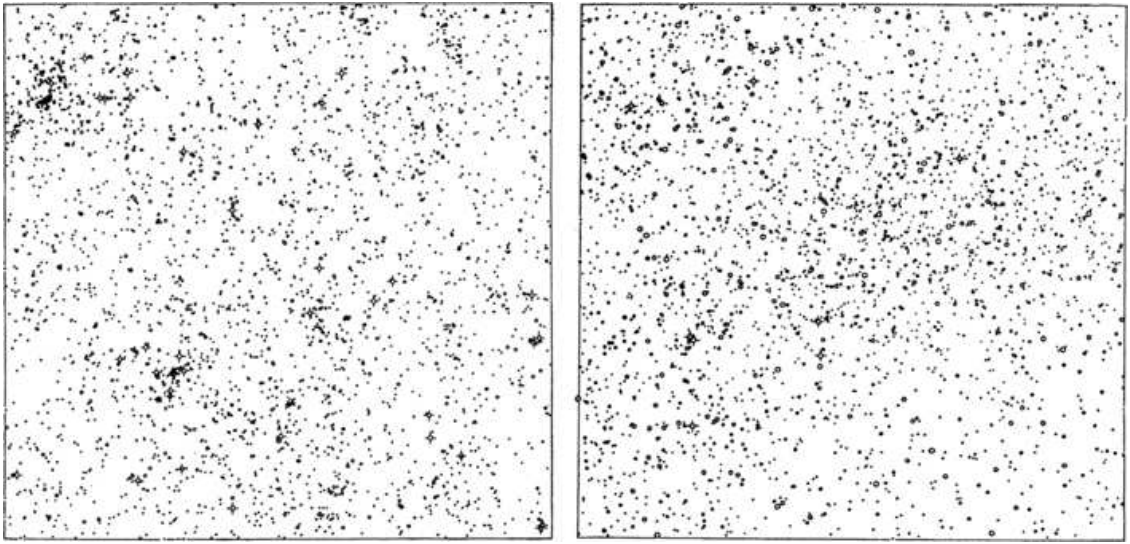

Fig. 1. *Left-hand panel is an image of an actual photographic plate. The right-hand panel is a synthetic plate. See Scott, Shane and Wirtanen ([1954](#)).*

In a series of papers Neyman, Scott, Shane and Swanson addressed the issue of galaxy clustering. They applied mathematical models to the Lick galaxy counts of Shane and Wirtanen. They were the first to compare the observed galaxy distribution to synthetic images of the Universe. They assumed that clusters occur around centers distributed as a spatial Poisson process. Each center was assigned a random number of galaxies and the latter placed independently at random distances from the center. This model, the so-called Neyman–Scott model, seemed to fit reasonably. However, when Neyman and Scott produced a simulated realization, or synthetic plate, of the sky from their model they were surprised. The actual pictures of the sky were a lot more lumpy than those their simulation had produced.

> "When the calculated scheme of distribution was compared with the actual distribution of galaxies ..., it became apparent that the simple mechanism postulated could not produce a distribution resembling the one we see" (Neyman and Scott, 1956).

More clustering was needed in the model. Neyman and Scott proceeded to introduce it. With a two-stage clustering process the simulated appearance of the sky looked much more realistic. Figure 1, taken from Scott, Shane and Wirtanen (1954), presents an example.

In summary,

> "... it was shown that the visual appearance of a 'synthetic' photographic plate, obtained by means of a large-scale sampling experiment, conforming exactly with the assumptions of the theory, is very similar to that of an actual plate" (Neyman, Scott and Shane 1954).

### 4.2 Example 2. SARDINE DEPLETION

> "Biometry is an interdisciplinary domain aimed at the understanding of biological phenomena in terms of chance mechanisms." (Neyman, 1976).

In 1947–1948 Neyman was called upon by the California Council of the Congress of Industrial Organizations to study the decrease in sardine catches. The decrease was of great concern and strongly affected the canneries and commerce of the workers along the west coast of the United States.

In particular JN was consulted regarding the natural and fishing mortality of the sardines. A specific purpose of his work was "... to study the methods of estimating the death rates of the sardines." JN wrote three reports on sardine fishery. They are collected in Neyman (1948) and titled, 1. *Evaluations and Observations of Material and Data Available on the Sardine Fishery*, 2. *Natural and Fishing Mortality of the Sardines,* and 3. *Contribution to the Problem of Estimating Populations of Fish with Particular Reference to Fish Caught in Schools, Such as*

*Sardines.* A revision of the third report appeared as Neyman (1949).

At the outset of Neyman (1949), he provides Table 2. From it he infers a "rapid decline ... observed in spite of a reported increase in fishing effort..." A second table, Table 3, gives the amount (in arbitrary units) of sardines landed on the West Coast in the seasons 1941–1946, classified by age and season. Figure 2 graphs the amounts with lines joining the values for the same sardine age. One sees the high numbers in the early 1940s followed by decline. The interpretation is tricky because the numbers reflect both the fish available and the effort put into catching them. Neyman (1948) discussed the effect of migration and concluded that it was unimportant for his current purposes.

Turning to analysis Neyman remarks,

> "Certain publications dealing with the survival rates of the sardines begin with the assumption that both the natural death rate and the fishing mortality are independent of the age of the sardines, at least beginning with a certain initial age." (Neyman, 1948).

and goes on to say,

> "In the present note a method is suggested whereby it is possible to a (sic) test the hypothesis that the natural death rate is independent of the age of the sardines" (Neyman, 1949).

To address the independence issue, and possibly motivated by Table 3, Neyman sets up a formal structure as follows. Let $N_{t,a}$ be "the number of fish available aged $a$ at the beginning of season $t$ and exposed to the risk of being caught." Here these numbers are collected into a vector, $\mathbf{N}(t) = [N_{t,a}]$. Next $n_{t,a}$ is set to be the expected number of sardines aged $a$ caught during season $t$, and $P_t = 1 - Q_t$ set to be the "fishing survival rate in the $t$th year." Continuing, $p_a = 1 - q_a$ denotes the "natural survival rate at age $a$" and $q_a$ the "rate of disappearance." The rate of mass emigration during season $t$ is denoted by $M_t$.

The following null hypothesis may be set down concerning the mortality rates,

$$H_0 : q_{a_0} = q_{a_0+1} = \cdots = q_a, \quad a > a_0.$$

Specific assumptions Neyman considered were:

(i) $Q_t = n_{t,a}/N_{t,a}$, season $t$ fishing mortality,
(ii) $N_{t+1,a+1} = N_{t,a}(1 - Q_t)(1 - q_a)$,
(iii) $N_{t+1,a+1} = N_{t,a}(1 - Q_t)(1 - M_t)(1 - q_a)$.

Assumptions (ii) and (iii) involve separation of the age and season variables. For identifiability of the model Neyman writes

$$n_{t+1,a+1} = n_{t,a} R_t p_a = n_{t,a} r_t p_a^*$$

with

$$R_t = \frac{P_t(1 - M_t)}{Q_t} Q_{t+1}, \quad r_t = R_t/R_1, \quad p_a^* = R_1 p_a.$$

TABLE 2
*Seasonal catch of California sardines 1943–1948 in 1000 tons*

| Year | Seasonal catch |
|---|---|
| 1943–1944 | 579 |
| 1944–1945 | 614 |
| 1945–1946 | 440 |
| 1946–1947 | 248 |
| 1947–1948 | 110 |

TABLE 3
*Numbers, $m_{t,a}$, of sardines caught by age and year*

| Season | 41–2 | 42–3 | 43–4 | 44–5 | 45–6 |
|---|---|---|---|---|---|
| age = 1 | 926.0 | 718.0 | 1030.0 | 951.0 | 493.0 |
| 2 | 6206.0 | 2512.0 | 1308.0 | 2481.0 | 1634.0 |
| 3 | 3207.0 | 4496.0 | 2245.0 | 1457.0 | 1529.0 |
| 4 | 868.0 | 1792.0 | 2688.0 | 1298.0 | 799.0 |
| 5 | 361.0 | 478.0 | 929.0 | 1368.0 | 407.0 |
| 6 | 95.1 | 169.4 | 327.0 | 498.5 | 299.2 |
| 7 | 47.2 | 36.0 | 98.4 | 148.0 | 111.2 |

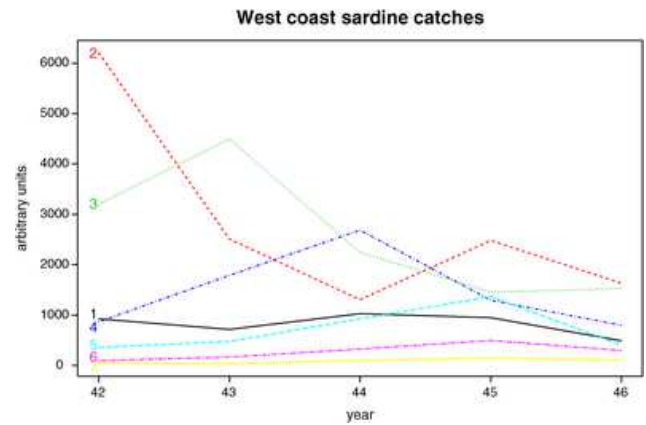

FIG. 2. *The data of Table 3 plotted versus year. The curve labels 1–7 index the age groups.*

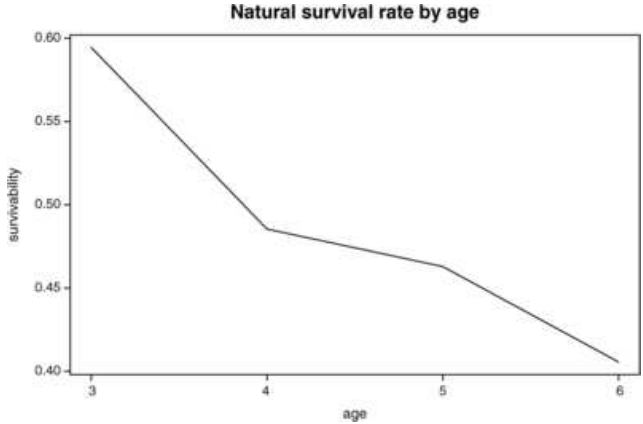

FIG. 3. *Estimates of the natural survival rate, $p^*$ as a function of age.*

One notes from these expressions that $n_{t+1,a+1}/n_{t,a}$ separates into a function of $t$ and a function of $a$. This last led Neyman to work with logs of ratios in his analyses. (There will be more on this choice later.) He estimates $p_a^* = R_1 p_a$, which is proportional to $p_a$ under his definitions, from the data.

The $p_a^*$ estimates are provided in Table 4 and graphed in Figure 3. One sees a steady decrease with age. Table 5 provides $\hat{n}_{t,a}$ based on assumptions (i) and (ii) [or (i) and (iii)].

Neyman's conclusions included,

> "While in certain instances the differences between Tables IV (here Table 3) and VII (here Table 5) are considerable, it will be recognized that the general character of variation in the figures of both tables is essentially similar" (Neyman, 1948, pages 14–15).

No formal test of $H_0$ was set down, but Neyman concludes that,

> "Since the *estimates* of the $p_a^*$ decrease rather regularly, it seems that the true natural survival rates must decrease with the increase in age..." (Neyman, 1948).

TABLE 4
*Parameter estimates (these are the values obtained in calculations for this article)*

| Season  | 41–2   | 42–3   | 43–4   | 44–5   |
|---------|--------|--------|--------|--------|
| $p_a^*$ | 0.5944 | 0.4854 | 0.4629 | 0.4056 |
| $r_t$   | 1.0    | 1.2252 | 1.0695 | 0.6259 |

Basic elements of this example include working with empirical data, noting the age and season structure explicitly, and working with a Markov-like setup. Interestingly Neyman talks of an expected value, but no full probability model is set down.

In part this example is meant to get the reader in the mood for an age-structured population analysis to appear later in the paper.

The final example taken from Neyman's work follows.

### 4.3 Example 3. WEATHER MODIFICATION

> "The meteorological aspects of planning an experiment with cloud seeding depend upon the past experience, upon what the experimenter is prepared to adopt as a working hypothesis and upon the questions that one wishes to have answered by the experiment" (Neyman and Scott, 1965–1966).

Cloud seeding became an interest of Jerzy Neyman starting in the early 1950s. He and his collaborators studied data from the Santa Barbara and Arizona rainfall experiments. Neyman and Scott moved on to study data from a Swiss weather modification experiment that had been designed to see if cloud seeding could reduce hailfall. The experiment was carried out in the Canton of Ticino during the period 1957–1963 and was called Grossversuch III.

The experimental design involved each day deciding whether conditions were suitable to define an "experimental day." If a day was suitable seeding was or was not carried out the following day, randomly. Seeding, if any, lasted from 0730 to 2130 hours local time. Rainfall measurements that had been made in Zurich, about 120 km away from Ticino, were studied.

In the course of their work Neyman and Scott discovered so-called "far-away effects," that is, an apparent increase in amount of rainfall at a distance. See Neyman, Scott and Wells (1969).

TABLE 5
*Estimates of the $n_{t,a}$, the expected numbers of sardines*

| Season | 1      | 2      | 3      | 4      | 5     |
|--------|--------|--------|--------|--------|-------|
| age, 3 | 2810.0 | 3556.3 | 2117.9 | 1761.6 | —     |
| 4      | 1059.3 | 1684.3 | 2611.7 | 1355.7 | 661.0 |
| 5      | 383.7  | 514.2  | 1001.7 | 1355.7 | 412.5 |
| 6      | 91.9   | 77.6   | 291.6  | 495.9  | 391.7 |
| 7      | —      | 37.3   | 88.2   | 126.5  | 125.9 |

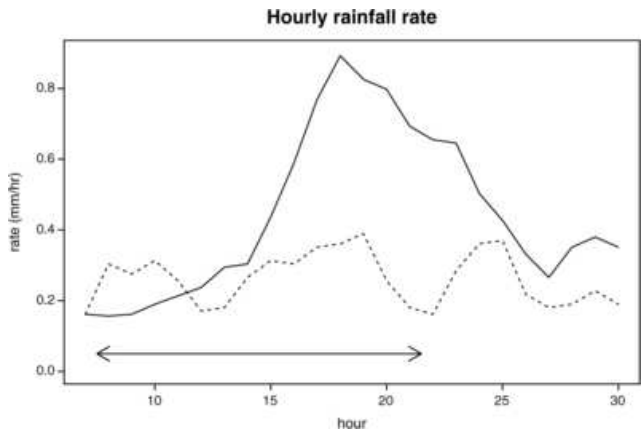

Fig. 4. *Comparison of seeded and not seeded hourly precipitation amounts on days with southerly upper winds. The solid line is rainfall for seeded days and the dashed line for unseeded. The horizontal line with arrowheads represents the seeding period at Ticino. A three-hour moving average had been employed to smooth hourly totals.*

Figure 4 provides a reconstruction of a graph that Neyman and Scott (1974) employed to highlight the result. It presents average hourly rainfall totals smoothed by a running mean of 3, for the experimental days when a "warm" stability layer and southerly winds were present.

To obtain the data of Figure 4 the values were read off a graph in Neyman and Scott (1974). The solid curve refers to experimental days with seeding, the dashed to those without. There were 53 experimental days with seeding and 38 without.

What Neyman and Scott focused on in the figure was an apparent effect of seeding in Zurich starting about 1400 hours in the afternoon.

They wrote as follows,

> "...the curves...represent averages of a number of independent realizations of certain stochastic processes. The 'seeded' curves are a sample from a population of one kind of processes and the 'not seeded' curve a sample from another. For an initial period of a number of hours...the two kinds of processes coincide. Thereafter, at some unknown time T, the two processes may become different. Presumably, all the experimental days differ from each other, possibly depending on the direction and velocity of prevailing winds. Therefore, the time T must be considered as a random variable with some unknown distribution. *The theoretical problem is to deduce the confidence interval for the expectation of T,...*" (Neyman and Scott, 1974).

This problem will be returned to later in the paper.

### 4.4 Neyman and Exploratory Data Analysis (EDA)

Given my statistical background it would be remiss not to provide some discussion of EDA in Neyman's work. Quotes are one way to bring out pertinent aspects of Neyman's attitude to EDA. One can conclude that exploratory data analysis was one of his talents.

> "...while hunting for a big problem I certainly established the habit, ..., to neglect rigour" (Neyman, 1967).

> "PAGE asked whether the elimination of outliers–supposed projected foreground or background objects recognized by discordant velocities–would not in itself introduce unwanted selection effects. NEYMAN advised that the investigator try calculations with and without outliers, then make up his mind 'which he likes best', while retaining both."

> "Compared with the old style experiments, characterized by the attitude 'to prove,' the proposed experiment would be substantially richer.... This, then, will implement the attitude 'to explore' contrasted with that 'to prove'" (Neyman and Scott, 1965–1966).

> "We emphasize that such an investigation is only exploratory; whatever may be found are only clues which must be studied further and hopefully verified in other experiments" (Dawkins, Neyman and Scott, 1977).

JN did not seem to use residuals much. However, in Neyman (1980) one does find,

> "...one can observe a substantial number of consecutive differences that are all negative while all the others are positive. ...the 'goodness of fit' is subject to a rather strong doubt, irrespective of the actual computed value of $\chi^2$, even if it happens to be small" (Neyman, 1980).

Neyman et al. (1953) proposed an innovative EDA method to examine variability: specifically, given values $X$ and $Y$ with the same units, plot $X - Y$ and $|X - Y|$ versus $(X + Y)/2$. Figure 5 compares Tables 3 and 5 of the sardine analysis this way. In the two panels one sees wedging, that is, an increase of variability with size. This suggests that a transformation of the data might simplify the matter. Neyman did employ the log transform in his analysis of the sardine data consistent with the multiplicative character of the model.

## 5. FOUR EXAMPLES OF RANDOM PROCESS DATA ANALYSIS

The following examples report some of my work, typically with collaborators. They were suggested in part by my exposure to JN and to the preceding examples.

### 5.1 Example 4. SHEEP BLOWFLIES

In Example 2 above Neyman studied data on sardines that included the actual age information. However, it can be the case that, even though a population is age-structured, only aggregate data are available, and actual age information is unavailable. This is the case in the example that follows. To deal with it a state-space model is set down. The (unobserved) state vector is taken to be the counts of individuals in the various age groups. The story and details follow.

The tale begins with the mathematician John Guckenheimer and the then entomologist George Oster coming to meet with DRB. They had in hand data on a population of *lucilia cuprina* (Australian sheep blowflies). The data concerned an experiment maintained from 1954 to 1956 under constant, but limited conditions by A. J. Nicholson, then Chief Division of Entomology, CSIRO, Australia.

At the beginning of the experiment 1000 eggs were placed in a cage. Every other day counts were made of the number of eggs, of nonemerging flies' eggs, of the number of adult flies emerging, and of the number of adult fly deaths. The life stages, and corresponding time periods, of these insects are given in Table 6. Further details of the experiment may be found in Nicholson (1957). To get digital values Oster and a student took a photo of one of the figures in that paper. The photo was then projected on a wall and numerical values read off. Unfortunately some of the populations' sizes went off the top of the figure. The values for these cases were obtained when DRB later visited CSIRO.

Guckenheimer and Oster's question was whether these data displayed the presence of a strange attractor, a concept from nonlinear dynamic systems analysis; see Brillinger et al. (1980) and Guckenheimer and Holmes (1983). The behavior evidenced in the second half of the series graphed in Figure 6 is what attracted Guckenheimer and Oster's attention. The initial oscillations come from the usual lifespan of the adults.

In the particular experiment studied here the amount of food put in the fly cage was deliberately restricted. This meant that the fecundity of the females was reduced. When much food was available many eggs were laid. With insufficient food the number of eggs was reduced. This led to boom periods and bust periods in the population size.

Figure 6 graphs the square roots of total adult population count, as well as of the number of flies emerging. The time points are every other day over a period of approximately two years. In the graphs one sees an initial periodic behavior in both series followed by rather irregular behavior. The square roots

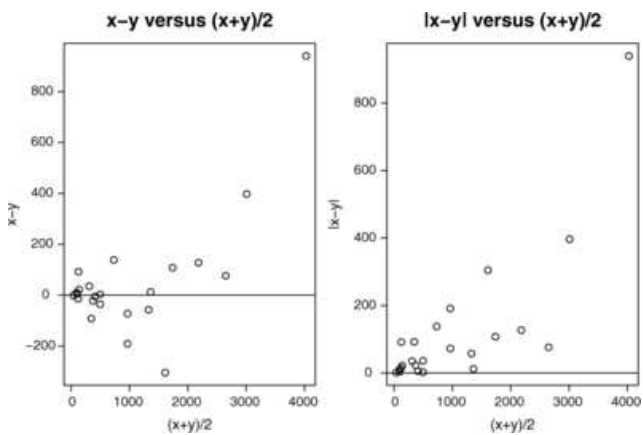

FIG. 5. *Comparisons of Table 3, x-values and Table 5, y–values. The left panel plots $(x - y)$ versus $(x + y)/2$ and the right $|x - y|$ versus $(x + y)/2$.*

TABLE 6
*Life stages and their lengths for sheep blowflies*

| Life stage | Length |
| --- | --- |
| egg | 12–24 hours |
| larva | 5–10 days |
| pupa | 6–8 days |
| adult | 1–35 days |

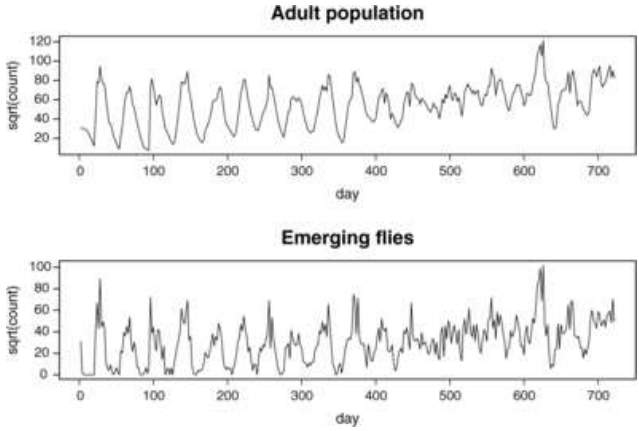

Fig. 6. *Square roots of counts for the Nicholson blowfly data. The top panel provides the number of adults and the bottom the number of emerging pupae.*

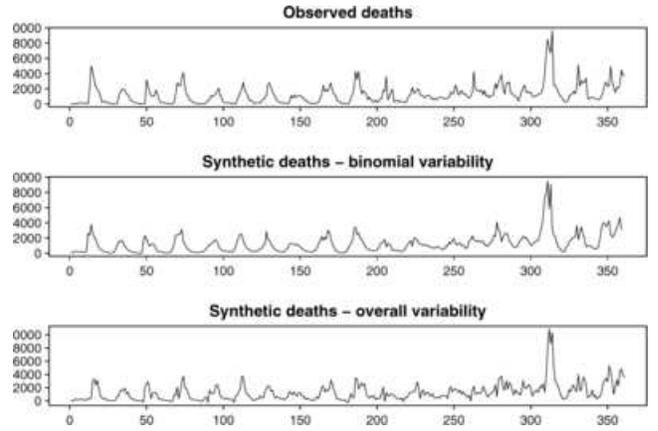

Fig. 7. *Death series and synthetic death series using the model* (6).

were plotted to make the variability of the display more nearly constant.

Brillinger et al. (1981) proceeded by setting down a formal state-space model for the situation as follows:

$t = 0, 1, 2, \ldots$, represents time, observations being made every other day,

$E_t$, the number of emerging flies in time period $(t, t+1]$,

$\mathbf{E}_t$, the entrant column vector; it has $E_t$ in row 1 and 0 elsewhere,

$N_t$, the adult population at time $t$.

Constructs include:

$\mathbf{N}_t = [N_{it}]$, the state vector; in it row $i$ gives the number of population members aged $i-1$ at time $t$,

$\mathbf{P}_t = \mathbf{P}(\mathbf{H}_t) = [p_{i,t}]$, the survival matrix. The entry in row $i+1$, column $i$ gives the proportion surviving age $i$ to age $i+1$. $\mathbf{P}_t$ is taken as depending on the history $\mathbf{H}_t$, that is, the collection of the data values up to and including time $t$.

The available data are $E_t$ and $N_t$.

The measurement equation, corresponding to the observed population size is, $N_t = \mathbf{1}'\mathbf{N}_t$. The dynamic equation is

$$\mathbf{N}_{t+1} = \mathbf{P_t N_t} + \mathbf{E}_{t+1} + \textit{fluctuations}.$$

This expression updates the counts of adult flies in each age group, starting from $\mathbf{N}_0 = 0$. The fluctuations represent variabilities in those numbers.

In one analysis (Brillinger et al., 1980), the following nonlinear age and density model was employed:

$p_{i,t} = 1 - \text{Prob}\{\textit{individual aged } i,$

$\textit{dies aged } i \textit{ at time } t | \mathbf{H}_t\}$

(6)      $= (1 - \alpha_i)(1 - \beta N_t)(1 - \gamma N_{t-1}).$

This model allows survival dependence on age, $i$, on the current population size, $N_t$ and on the preceding population size, $N_{t-1}$. The final term allows the possibility that it takes some time for the limited or excess food situation to take effect.

Weighted least squares was employed in the fitting of model (6). On the basis of residual plots weights were taken to be $N_t^2$. Hence writing $D_t = N_{t-1} - N_t + E_t$ one seeks

$$\min_\theta \sum_t \left( D_{t+1} - \sum_i q_i m_{i,t} \right)^2 \Big/ N_t^2,$$

where $\theta = \{\alpha_i, \beta, \gamma\}$ and $m_{i,t}$ is the conditional expected value, $E\{N_{i,t}|\mathbf{H}_t\}$. Graphs of the estimates of the individual entries of $\mathbf{N}_t$ are provided in Brillinger et al. (1980).

Synthetic series were computed to assess the reasonableness of the model (6). In the simulations counts of deaths in the time period $(t-1, t]$, are computed. The deaths, $D_t$, are plotted in the top panel of Figure 7. The value $D_t$ is thought of as fluctuating about the value

$$\sum_i q_{i,t} N_{i,t}$$

where $N_{i,t}$ is the population aged $i$ at time $t$.

The results of two simulations are provided in Figure 7. In the first, the middle series, the variability is taken as binomial. In the second, the bottom series, the variability is taken as independent normal, mean 0, standard deviation $\hat{\sigma} N_t$ with $\hat{\sigma}$ estimated from the weighted least squares results. That the appearances of the synthetic series are so close to the actual series relates to the use of the common stimulus series, $E_t$.

A byproduct of this analysis is that because the measurement equation, $N_t = \mathbf{1}'\mathbf{N}_t$, is of simple addition form by this analysis one has developed a decomposition of the population total series into individual age series. These are graphed in Brillinger et al. (1980).

The fitted death rates were nonlinear in the population size, so mathematically a strange attractor might be present (Brillinger, 1981).

In this situation one is actually dealing with a nonlinear closed loop feedback system with time lags. Guttorp (1980), in his doctoral thesis, completed the analysis of the feedback loop modeling the births.

### 5.2 Example 5. WEATHER MODIFICATION REVISITED

Neyman and Scott's problem referred to in Example 3 was addressed in Brillinger (1995). At issue was making inferences concerning the travel time of seeding effects from Ticino to Zurich. The approach of the paper was to envisage a succession of travel time effects that started at times throughout the seeding period. This way one had replicates to allow employment of statistical characteristics. A conceptual model involving a gamma density for the travel velocity of the seeding effect was employed. The data themselves were graphed in Figure 4 above.

The model employed is the following. Suppose that "rain particles" created at Ticino move off toward Zurich with a possibility of leading to a cluster of rain drops there. Suppose that the particles are born at Ticino at the times $\sigma_j$ of a point process $M$, at rate $p_M(t)$. Suppose that the travel times from the particles' times of creation, $U_j$, to Zurich are independent of each other with density $f_U(\cdot)$. Let $N$ denote the point process of times, $\tau_j$, at which the particles arrive at Zurich and $p_N(t)$ denote the rate of that process.

If the $j$th particle moves with velocity $v_j$ and the distance to be traveled is $\Delta$, then its travel time is $u_j = \Delta/v_j$ and since

$$\sum_j \delta(t - \tau_j) = \sum_j \delta(t - \sigma_j - u_j)$$

with $\delta(\cdot)$ the Dirac delta, one has

$$p_N(t) = \int p_M(t-u) f_U(u)\, du.$$

Let the amounts, $R_j$, of rain falling at Zurich associated with the individual particles, be statistically independent of the particles. Let $\mu_R$ denote $E\{R_j\}$. Then the rate of rainfall at Zurich at time $t$ is

$$p_X(t) = \mu_R \int p_M(t-u) f_U(u)\, du.$$

Next let $X(t)$ denote the cumulative amount of rain falling at Zurich from time 0 to time $t$. Its expected value is

$$E\{X(t)\} = \int_0^t p_X(v)\, dv.$$

Turning to Figure 4, Neyman and Scott employed a running mean of order 3 of the hourly totals to get the values graphed. These are the data available for analysis. (The hourly values appear to be lost.) The running mean may be written

$$Y(t) = \tfrac{1}{3}(X(t+1) - X(t-2))$$

for $t = 2, 3, \ldots$. Its expected value is

$$
\begin{aligned}
&\tfrac{1}{3}\int_{t-2}^{t+1} p_X(v)\, dv \\
&\qquad = \tfrac{1}{3}\mu_R \int_{t-2}^{t+1} \int p_M(v-u) f_U(u)\, du\, dv.
\end{aligned}
\tag{7}
$$

One can now view the Neyman–Scott problem as related to estimating $f_U(\cdot)$ of (7), that is, estimating the travel time density given the available data.

To proceed, the seeding rate $p_M(t)$ will be taken to be constant on the time interval from 0730 to 2130 hours and to be 0 otherwise. It will be further assumed that the travel time of $U$ has the form $\theta/W$ with $\theta$ a parameter, and with $W$ Weibull, having scale 1, and shape $s$. Brillinger (1995) took the gamma as the density, but a review of the literature of wind speeds suggests that the Weibull would be more appropriate.

Writing $p_M(t) = C$ for $A < t < B$ (here A = 7.5 and B = 21.5 hr) one has the regression function

$$
\begin{aligned}
E\{Y(t)\} = \alpha + \frac{C}{3}\mu_R \bigg[&\int_{t-2-A}^{t+1-A} F_U(u)\, du \\
&- \int_{t-2-B}^{t+1-B} F_U(u)\, du \bigg],
\end{aligned}
\tag{8}
$$

where $F_U(\cdot)$ denotes the distribution function of $U$, in the case of seeding and $\alpha$ is the natural level of rainfall. With the assumed Weibull velocity distribution, (8) may be evaluated in terms of $G$ the distribution function of the Weibull. Specifically,

$$
\begin{aligned}
\int_0^x F_U(u)\, du = &\, x\left[1 - G\left(\tfrac{1}{x}, s\right)\right] \\
&- \frac{s}{s-1}\left[1 - G\left(\tfrac{1}{x}, s-1\right)\right].
\end{aligned}
$$

(To derive this one replaces Prob$\{1/W \leq u\}$ by Prob$\{W \geq 1/u\}$ and integrates by parts.)

The estimates of the unknowns $\mu = \theta\Gamma((s-1)/s)$ (the average travel time), $s$, $\alpha$, $\beta = C\mu_R/3$ were determined by ordinary least squares, weighting the seeded terms by 53 and the unseeded by 38 to handle the unequal numbers of seeded and unseeded cases.

Figure 8, left-hand panel, presents the data (solid curve) and the fitted (dotted) curve. The parameter estimates obtained are:

$$\hat{\mu} = 4.78(0.47) \text{ hr},$$
$$\hat{s} = 6.68(5.12),$$
$$\hat{\alpha} = 0.24(0.02),$$
$$\hat{\beta} = 1.69(0.19).$$

[The standard errors, assumed the errors to be i.i.d.]

One sees in the left-hand panel that the actual data have a peak near 1800 during 0730 and 2130 hr, whereas the fitted has a flat top. Perhaps the birthrate, $p_M(t)$, of particles is not approximately constant as assumed above. Perhaps the distribution, $f_U(u)$, depends on time. Perhaps the result is due to natural variability.

A synthetic plot is generated to examine the fit. Specifically the fluctuations of the unseeded days have been added to the fitted curve and graphed in the right-hand panel of Figure 8. Still the fitted curve is quite flat on the top, in contrast to the Neyman–Scott data curve which is noticeably peaked. The added fluctuations do not bring the curve up to the data level.

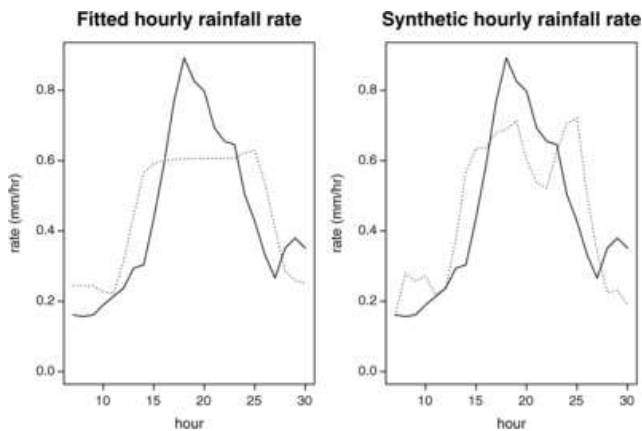

FIG. 8. *Left panel—actual and fitted (dotted line) rainfall when seeding. Right panel—actual and synthetic in the case of seeding (dotted line).*

Returning to the Neyman–Scott problem of Section 3, the second quotation there refers to $T$, a random time at which seeding first shows up in Zurich. The $U$'s represent the lengths of time it takes for an effect just initiated to arrive. One can take the expected value, $EU$, to be $ET$. Using the parameter estimates above, an approximate 95% confidence interval for the expectation of $T$ is

$$4.78 \pm 2 * 0.47 \text{ hours}.$$

More work needs to be done with this example. A indication of how to proceed is provided by Figure 8. The data graph is pointed, whereas the fitted is flat-topped.

### 5.3 Example 6. ELK MOTION

The data now studied were collected at the Starkey Experimental Forest and Range (Starkey), in Northeastern Oregon. Quoting from the website, fs.fed.us/pnw/starkey/publications/by_keyword/Modelling_Pubs.shtml.

Starkey was set up by the US Forest Service for

> "Long-term studies of elk, deer, and cattle—examining the effects of ungulates on ecosystems."

A specific management question of concern is whether recreational uses by humans would affect the animals there substantially. Further details about Starkey and the recreation experiment may be found in Brillinger et al. (2001a, 2001b, 2004), Preisler et al. (2004) and Wisdom (2005).

In the first analysis presented the elk were not deliberately disturbed and their paths were sampled at discrete times. This gave control data for an experiment. An all-terrain vehicle (ATV) was introduced and driven around on the roads in the NE Meadow of Starkey. The analysis to be presented quantifies the effect of the disturbance. The locations of both the ATV and the elk were monitored by GPS methods.

There were 8 elk in the study. The ATV was introduced into the meadow over 5-day periods. This was followed by 9-day "control" periods with no ATV. In the control periods the animals were located every 2 hours. In the ATV case elk locations were estimated about every 5 min. The ATV's locations were determined every second.

Figure 9, left-hand panel, shows observed elk trajectories superposed. One sees the animals constrained by the fence, but moving about most of the Reserve.

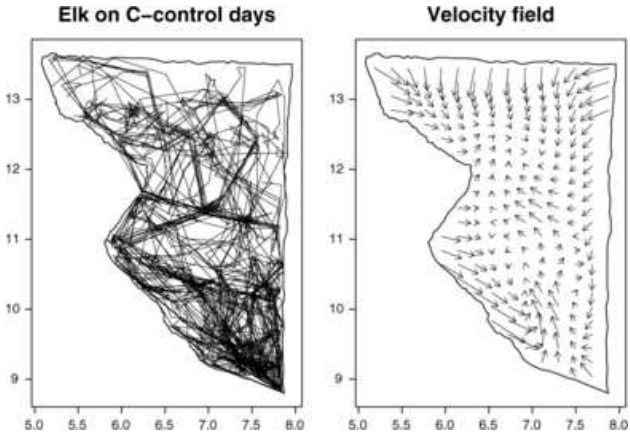

FIG. 9. *Northeast pasture of the Starkey Reserve and the elk motion on control days. The left panel shows the paths of 8 elk, superposed. The right panel displays the estimated velocity field $\hat{\boldsymbol{\mu}}(\mathbf{r})$ as a vector field.*

They often visit the SE corner. The straight line segments result from the locations being obtained only every 2 hours in this control case.

The animal motion will be modeled by the SDE

$$(9) \qquad d\mathbf{r}(t) = \boldsymbol{\mu}(\mathbf{r}(t))\,dt + \boldsymbol{\sigma}\,d\mathbf{B}(t)$$

with $\mathbf{r}(t)$ the location at time $t$, $\mathbf{B}$ a bivariate standard Brownian motion and $\sigma$ a scalar. The function $\boldsymbol{\mu}$ is assumed to be smooth. The discrete approximation (3) becomes a generalized additive model with Gaussian errors; see Hastie and Tibshirani (1990).

The resulting estimate is displayed as a velocity vector field $(\hat{\mu}_1(\mathbf{r}), \hat{\mu}_2(\mathbf{r}))$ in the right-hand panel of Figure 9 employing arrows. One sees the animals moving along the boundary and toward the center of the pasture. The fence can be ignored in this data analysis.

The fence is important in preparing a synthetic trajectory. What was done in that connection was to employ the relation (3) with the proviso that if it generated a point outside the boundary, then another point was generated until one stayed within the boundary. This is a naive but effective method if the $t_i$ of (3) are close enough together. Better ways for dealing with boundaries are reviewed in Brillinger (2003).

Figure 10 shows the trajectories of three of the animals. The lower right panel presents a synthetic path generated including 188 location points. The synthetic trajectory does not appear unreasonable.

Consideration now turns an analog of regression analysis for trajectories, that is, there is an explanatory variable. The explanatory variable is the changing location, $\mathbf{x}(t)$, of the ATV. The left-hand panel of Figure 11 shows the routes of the ATV cruising around the roads of the Meadow. The right-hand panel provides the superposed trajectories of the 8 elk. One sees, for example, the elk heading to the NE corner, possibly seeking refuge. The noise of the ATV is surely a repellor when it is close to an elk, but one wonders at what distance does the repulsion begin?

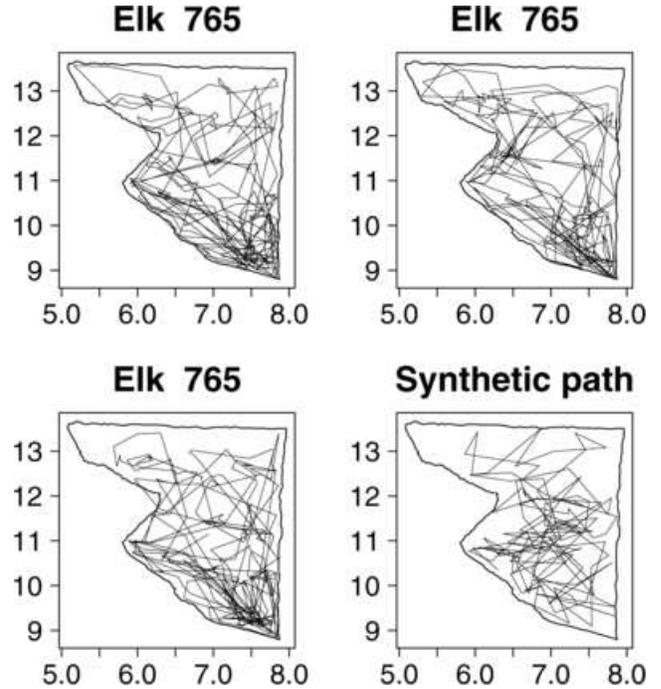

FIG. 10. *The first three panels display the tracks of the indicated animals. The final panel, lower right, is a synthetic path.*

The following model was employed to study that question. Let $\mathbf{r}(t)$ denote the location of an elk, and $\mathbf{x}(t)$ the location of the ATV, both at time $t$. Let $\tau$

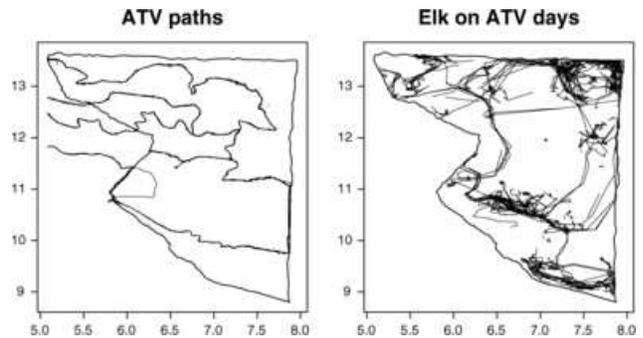

FIG. 11. *The left panel shows the ATV's route, while the right shows the elk paths in the presence of the ATV. The ATV passes in and out some gates on the lefthand side.*

be a time lag to be studied. Consider

$$dr(t) = \mu(r(t))\,dt + \nu(|r(t) - x(t-\tau)|)\,dt \quad (10)$$
$$+ \sigma\,dB(t).$$

The times of observation differ for the elk and the ATV. They are every 5 minutes for the elk when the ATV is present and every 1 sec for the ATV itself. In the approach adopted location values, $x(t)$, of the ATV are estimated for the elk observation times via interpolation. The ATV observed times are close in time, namely 1 second, so the interpolation should be reasonably accurate.

Expression (10) allows the change in speed of an elk to be affected by the location of the ATV $\tau$ time units earlier. Assuming that $\mu$ and $\nu$ in (10) are smooth functions, then the model may be fit as a generalized additive model. Figure 12 graphs $|\hat{\nu}(d)|$, $d$ being the distance of the elk from the ATV. (The norm $|\nu| = \sqrt{\nu_1^2 + \nu_2^2}$ here.) One sees an apparent increase in the speed of the elk, particularly when an elk and the ATV are close to each another. The increased speed is apparent at distances out to about 1.5 km. An upper 95% null level is indicated in Figure 12 by a dashed line. One sees less precise measurement at increasing large values of $\tau$.

The estimation of $|\nu(d)|$ was also carried out in the absence of the $\mu$ term in the model. The results were very similar. This gives some validity to interpreting the estimate $\hat{\nu}(d)$ on its own despite the presence of $\mu$ in the model.

In conclusion, the ATV is having an apparent effect and it has been quantified to an extent by the graphs of Figure 12.

These results were presented in Brillinger et al. (2004). Also Wisdom (2005) and Preisler et al. (2004) modeled the probability of elk response to ATVs in a different way. They used data for the year 2002, and measured the presence of an effect in another manner.

### 5.4 Example 7. MONK SEALS: A POTENTIAL FUNCTION APPROACH

Hawaiian monk seals are endemic to the Hawaiian Islands. The species is endangered and has been declining for several decades. It now numbers about 1300. One hypothesis accounting for the decline in numbers is the poor growth and survival of young seals owing to poor foraging success. In consequence of the decline data have been collected recently on the foraging habitats, movements, and behaviors of these seals throughout the Hawaiian Islands Archipelago. Specific questions that have been posed regarding the species include:

What are the geographic and vertical marine habitats that Hawaiian monk seals use?

How long is a foraging trip?

For more biological detail see Stewart et al. (2006) and Brillinger, Stewart and Litnnan (2006, 2008).

The data set studied is for the west side of the main Hawaiian Island of Molokai. The work proceeds by fitting an SDE that mimics some aspects of the behavior of seals. It employs GPS location data collected for one seal. An SDE is found by developing a potential function.

The data are from a three-month journey of a juvenile male while he foraged and occasionally hauled out onshore. The track started 13 April 2004 and ended 27 July 2004. The animal was tagged and released at the southwest corner of Molokai; see Figure 13, top left panel. The track is indicated for six contiguous 15-day periods. The seal had a satellite-linked radio transmitter glued to his dorsal pelage. It was used to document geographic and vertical movements as proxies of foraging behavior.

There were 754 location estimates provided by the Argos satellite service, but many were suspicious. Associated with a location estimate is a prediction of the location's error (LC or location class). The LC index takes on the values 3, 2, 1, 0, A, B, Z. When LC = 3, 2 or 1 the error in the location is predicted to be 1 km or less, and these are the cases employed in the analysis here.

The estimated times of locations are irregularly spaced and not as close together as one might like. This can lead to difficulties of analysis and interpretation.

The motivating SDE of the analysis is

$$(11) \quad dr(t) = \mu(r(t))\,dt + \sigma\,dB(t), \quad r(t) \in F,$$

with $\mu = -\nabla H$, $H$ a potential function, $\sigma$ scalar, $B$ bivariate Brownian and $F$ the region inside the 200-fathom line up to Molokai. There was discussion of the potential approach in Section 3. The potential function employed here is

$$H(x,y) = \beta_{10}x + \beta_{01}y + \beta_{20}x^2 \quad (12)$$
$$+ \beta_{11}xy + \beta_{02}y^2 + C/d_M(x,y)$$

where $d_M$ is the shortest distance to Molokai from the location $(x,y)$. The final term in (12) is meant to keep the animal off Molokai.

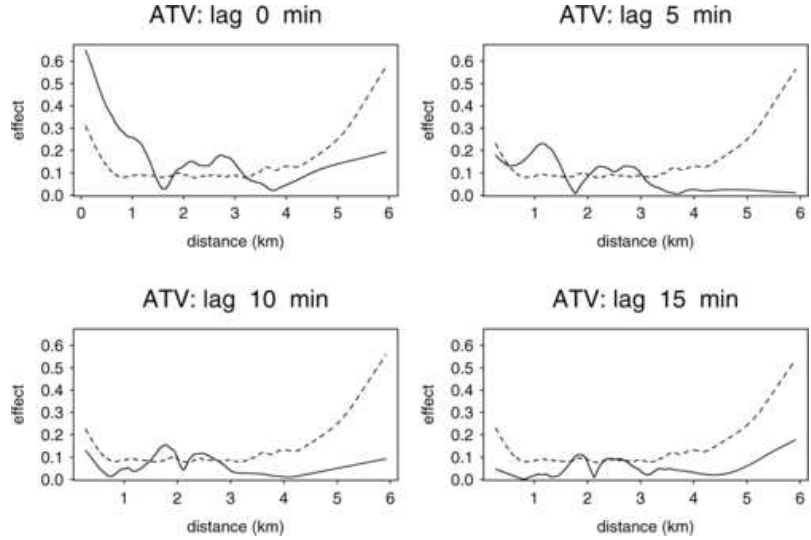

Fig. 12. *The function $|\hat{\boldsymbol{\nu}}|$ of (10) for the time lags 0, 5, 10, 15 minutes.*

The model was fit by ordinary least squares taking $C = 7.5$. In the analysis the number of data points was 142 and the parameter estimates obtained were $\hat{\boldsymbol{\beta}} = (93.53, 8.00, -0.47, 0.47, -0.41)$, and $\hat{\sigma} = 4.64$ km. Figure 14 shows the estimated potential function, $\hat{H}$. This seal is pulled into the middle of the concentric contours, with the Brownian term pushing it about.

Synthetic plots were generated to assess the reasonableness of the model and to suggest departures. Figure 15 shows the results of a simulation of the process (only one path was generated) having taken the parameter values to be those estimated and having broken the overall trajectory down into six segments as in Figure 13, to which it may be compared.

The sampling interval, $dt$, employed in the numerical integration of the fitted SDE is 1 hour. The paths were constrained to not go outside the 200-fathom line and not to go on the island. (See Brillinger, 2003, for methods of doing this.) The locations of the time points of the synthetic track are the times of the observed locations. This allows direct compar-ison with the data plot of Figure 13. The variability of Figure 15 is not unlike that of Figure 13.

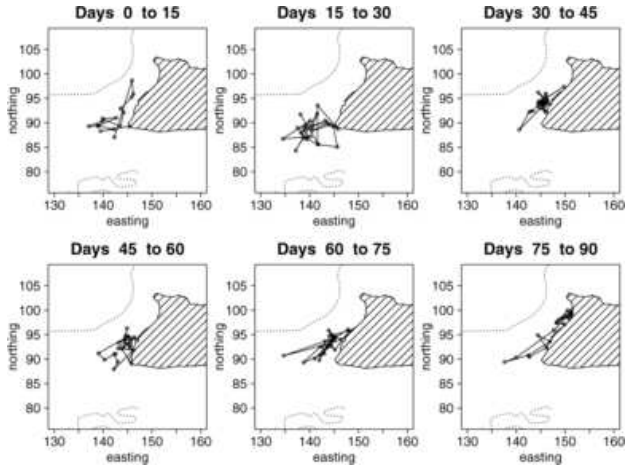

Fig. 13. *Plots of the seal's well-determined locations for successive 15-day periods. The dashed line is the 200-fathom line. It corresponds to Penguin Bank.*

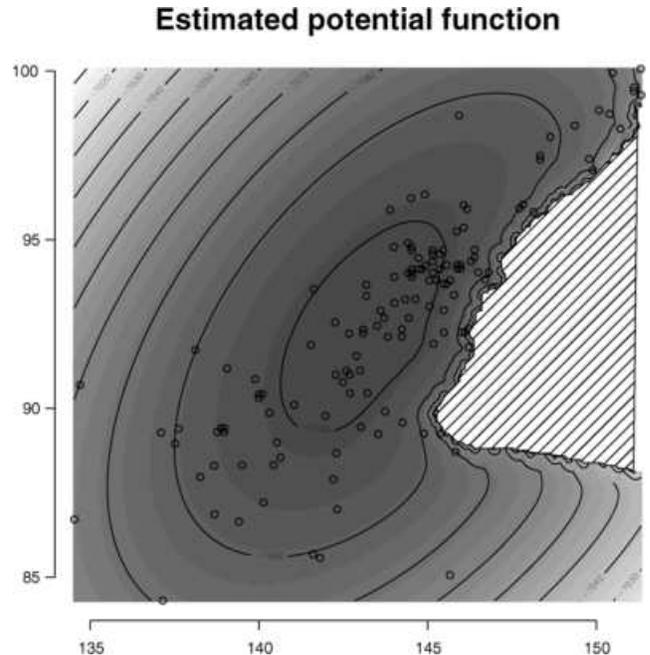

Fig. 14. *The fitted potential function obtained using the potential function (12). The darker the values are, the deeper the potential function is. The slanted line region is Molokai.*

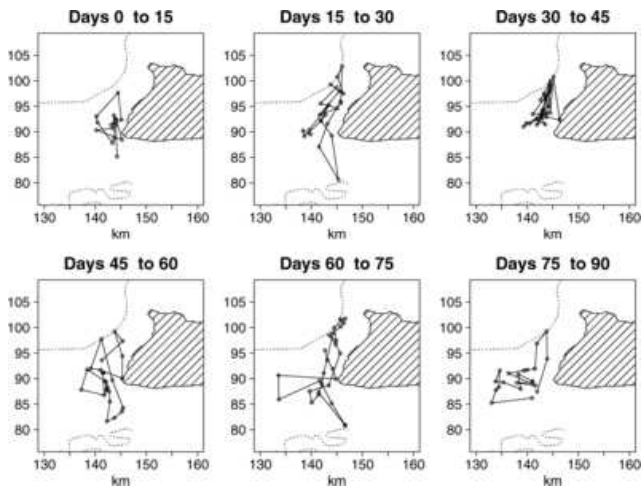

Fig. 15. *Synthetic plots of the model* (11) *having fit the potential function* (12). *The times are those of the data of Figure* 13.

In this work the scattered, sometimes unreasonable, satellite locations have been cleaned up and summarized by a potential function. The general motion of the animal on a foraging trip has been inferred and simulated. It has been learned that the animal stays mostly within Penguin Bank and tends to remain in an area off the west coast of Molokai.

There are other examples of potential function estimation in Brillinger, Stewart and Littnnan (2006, 2008) and Brillinger (2007a, 2007b).

## 6. CONCLUSION

> "Say what you are going to say, say it, then say what you said" (Neyman, Personal communication).

It was a great honor to be invited to present the Neyman Lecture. I attended many Neyman Seminars and made quite a few presentations as well. A side effect of the work was the very pleasant experience of reading through many of Neyman's papers in the course of preparing the lecture and the article. So many personal memories returned.

The emphasis has been placed on dynamic and spatial situations. There are three examples of JN and ELS; two concern temporal functions and one spatial. Four examples are provided of the work of DRB with collaborators. Two are temporal and two are spatial-temporal. The data are from astronomy, fisheries, meteorology, insect biology, animal biology and marine biology. The models and analyses were not all that difficult. The statistical package R was employed.

The field of sampling was another one to which Neyman made major contributions; see Neyman (1934, 1938a). It can be argued that work in sampling had a more profound impact on the United States than any of his other applied work. I looked hard but did not find reference to repeated sample surveys in JN's work. Had I, there would have been some discussion of dynamic sample survey.

The reader cannot have missed the many references to Elizabeth Scott. In fact in many places in my lecture the title could have been the Neyman–Scott Lecture. From the year 1948 on, 55 out of 140 of JN's papers were with her. Some 118 of Betty's publications are listed in Billard and Ferber (1991). One in the spirit of this lecture, Scott (1957), concerns the Scott effect, a biasing effect that occurs in galaxy observations because at greatest distances only the brightest would be observed. She developed a correction method (Scott, 1957).

I end with a wonderful and enlightening story concerning Jerzy Neyman. It was told by Alan Izenman at the lecture in Minneapolis. In the early 1970s the Berkeley Statistics Department voted to do away with language requirements. (There had been exams in two non-English languages.) In response in the graduate class that JN was teaching he announced that he was going to ask various people to give their presentation in their native, non-English, language. This continued for a number of weeks and languages.

## ACKNOWLEDGMENTS

I have to thank my collaborators and friends regarding the matters described in the paper. These include A. Ager, J. Guckenheimer, P. Guttorp, A. Izenman, J. Kie, C. Littnan, H. Preisler, G. Oster and B. Stewart. I also thank the Editor and referees for their comments. They improved the article. Brillinger (1983) was an earlier attempt at the topic.

The research was supported by various NSF grants, the most recent being DMS-05-04162, "Random processes: data analysis and theory."